# Rattling motion of a single atom in a fullerene cage molecule sensed by terahertz spectroscopy

Shaoqing Du[1, a)], Ya Zhang[1], Kenji Yoshida[1], and Kazuhiko Hirakawa[1,2,b)]

[1]*Institute of Industrial Science, University of Tokyo, 4-6-1 Komaba, Meguro-ku, Tokyo 153-8505, Japan*

[2]*Institute for Nano Quantum Information Electronics, University of Tokyo, 4-6-1 Komaba, Meguro-ku, Tokyo 153-8505, Japan*

Superatoms have been regarded as new building blocks for artificially synthesized electronic and magnetic materials and have been attracting great attention in nano-assembly and cluster science.[1,2]  Upon the discovery of the superatom states in fullerene[3] and its derivatives[4], particularly, endohedral metallofullerenes[5-7] (a positively charged core metal atom is surrounded by a negatively charged carbon cage), the superatom properties have become more attractive, because ultrafast motion of the trapped atom modifies the electron density distribution in a local area of a picometer-scale[8].  So far, the encapsulated atom position was imaged by scanning tunneling microscopy[9] (STM) or transmission electron microscopy[10] (TEM).  However, they can image the encapsulated metal atom only in a time-averaged manner; the observation of ultrafast atom motion is very challenging because dynamical processes take place in the terahertz (THz) frequency range[11,12] in a picometer-scale.  In the present work, a $C_{82}$ fullerene molecule that encapsulates a Ce atom is placed in a sub-nm gap of metal electrodes and excited by THz radiation to investigate the ultrafast dynamics of the Ce atom.  We have observed two THz-induced photocurrent peaks associated with the bending and stretching motions of the encapsulated Ce atom.  Furthermore, by measuring the bias voltage dependence of the THz-induced photocurrent, we have found that the THz-induced photocurrent flows not only by the photoconductive effect but also by the photovoltaic effect. This work demonstrates that THz spectroscopy using nanogap electrodes can detect a motion of a single atom, opening a door to ultrafast THz nanoscience at the sub-nm scale.

Observation of an ultrafast motion of an encapsulated single atom is very challenging due to a huge difference in size between the single atom and the THz wavelength[13-15]. One of the difficulties arises from the so-called "diffraction limit". Another difficulty is that, since the THz absorption by a single atom is extremely small, conventional absorption measurements cannot be used. Here, we report on the THz spectroscopy of an ultrafast motion of a single atom by using a single molecule transistor (SMT) geometry. The SMT is a structure in which a single molecule is captured in a sub-nm gap created between the source and drain metal electrodes fabricated on a field plate.[16-19] Figure 1a shows a schematic sample configuration used in this work. We used a $Ce@C_{82}$ endohedral metallofullerene (EMF) molecule, in which a single cerium (Ce) atom is encapsulated in a $C_{82}$ cage.[4-7,20] We captured a single $Ce@C_{82}$ molecule by the nanogap electrodes created by electromigration and fabricated $Ce@C_{82}$ SMTs. Using the source and drain electrodes as a THz antenna, we tightly focused THz radiation onto a $Ce@C_{82}$ molecule.[21,22] Furthermore, we have detected a very small THz absorption by a single atom by measuring THz-induced photocurrent.[22,23]

Figure 1b shows the scanning electron microscope (SEM) image of a sample used in this study. We deposited a 10-nm-thick NiCr layer on a high-resistivity Si substrate, which served as a semi-transparent backgate electrode. Then, a 30-nm-thick $Al_2O_3$ gate-insulation film was grown by atomic layer deposition. We fabricated gold nanojunctions on the surface of the wafer, using the shadow evaporation technique,[24-26] and formed the source and drain electrodes by applying the electrical break junction (EBJ) method to the nanojunctions. Details of the feedback control of the EBJ process are described elsewhere.[25] A bowtie antenna shape was used for the source and drain electrodes to achieve a good coupling efficiency between the THz radiation and the single $Ce@C_{82}$ molecule. As shown in Fig. 1c, the broad antenna resonance covers a frequency range from 0 to 5 THz (photon energy range $\hbar\omega$ = 0 - 20 meV). The electric-field distribution under the resonance condition ($\hbar\omega \sim 10$ meV) is plotted in the inset of Fig. 1c, showing that the THz field is concentrated in the nanogap region.[27] It was shown that the local THz field in the nanogap is enhanced by a factor of ~$10^5$ by the plasmonic effect of the metal electrodes.[21,27,28] The sample was glued on a hemispherical Si lens. A dilute toluene solution of Ce@$C_{82}$ molecules was deposited on the surface of the gold nanojunctions and dried off with nitrogen gas. The sample was then mounted in a vacuum space of a $^4$He cryostat. All the measurements were performed at ~4.6 K in a vacuum.

Figure 2a shows a color-coded differential conductance ($\partial I_{DS}/\partial V_{DS}$) map of a Ce@$C_{82}$ SMT plotted as a function of the source-drain voltage, $V_{DS}$, and the gate voltage, $V_G$ (Coulomb stability diagram). A crossing pattern in the diagram indicates that this device operates as a single electron transistor. Furthermore, vibrational excitation lines are observed at ~±5 meV, which suggests that a single Ce@$C_{82}$ molecule was trapped in the nanogap and served as a Coulomb island.[29] After the transport measurements, we illuminated the sample with a broadband THz radiation from a blackbody light source and searched for a THz-induced photocurrent. By sweeping $V_G$ while applying a very small $V_{DS}$ (= 0.1 mV), we measured the THz-induced photocurrent as a function of $V_G$. The black curve in Fig. 2b is the current measured in the dark condition, while the red trace is the THz-induced photocurrent. The peak of the black curve corresponds to the charge degeneracy point of the Ce@$C_{82}$ SMT. Due to thermal excitation, the Coulomb peak has a slightly asymmetric shape. The linear conductance at the Coulomb peak was 1.2 μS. Note that a small, but finite photocurrent of the order of 0.3 pA was observed near the charge degeneracy point.

Let us discuss how the photocurrent is generated in the Ce@$C_{82}$ SMT. When we have $N$ electrons on the molecule and the system is in the Coulomb blockaded region, as shown in Fig. 2c, the lowest unoccupied molecular orbital (LUMO) is above the source/drain Fermi levels. Therefore, an electron cannot enter the molecule in the dark condition. However, when the Ce@$C_{82}$ molecule is illuminated by THz radiation, an electron can hop onto the LUMO by absorbing a vibron (the vibron-assisted tunneling-in process). The electron excited to the LUMO level subsequently escapes the molecule to one of the electrodes, as shown in Fig. 2d. Similarly, when we have $N+1$ electrons on the molecule and the system is in the Coulomb blockaded region, the highest occupied molecular orbital (HOMO) is below the Fermi levels (see Fig. 2e) and the electron on the HOMO cannot escape the molecule in the dark condition. However, when the molecule is illuminated by THz radiation, the electron on the molecule absorbs a vibron and can escape the molecule (the vibron-assisted tunneling-out process). Subsequently, the emptied HOMO is refilled by an electron from one of the electrodes, as shown in Fig. 2f. This is consistent with the fact that the photocurrent measured by sweeping $V_G$ gradually decreases as the $V_G$ approaches the voltage where the extrapolated vibron excitation line intersects the $V_{DS} = 0$ line (see the dashed curve in Fig. 2b).

Note here that the photocurrent does not vanish even at $V_{DS}$ approaches 0, as seen in Fig. 2b, which implies that the THz photocurrent is partly generated by the photovoltaic effect. To confirm this, we plot in Supplementary Information the maps for the dark current and the photocurrent measured as a function of $V_G$ and $V_{DS}$ on a sample which exhibited a higher conductance (~ 10 μS) (see Fig. S1). The black dotted lines in Figs. S1c and S1d denote the boundaries of the Coulomb diamonds. The shot noise in the photocurrent is large in the single electron tunneling regions. In Fig. S1c, we can clearly see the boundaries of the Coulomb diamonds and see that the dark current changes its polarity, following the polarity of $V_{DS}$. In contrast, as seen in Fig. S1d, the THz-induced photocurrent does not depend on the bias voltage for the range plotted in the figure (- 1 mV < $V_{DS}$ < 1 mV), indicating that the photocurrent arises from the photovoltaic mechanism[30]. This photovoltaic effect originates from inherent asymmetry in the four tunneling rates, namely, $\Gamma_{Si}$, $\Gamma_{Di}$, $\Gamma_{So}$, and $\Gamma_{Do}$ (see Figs. 2d & 2f).[30] Here, the subscripts S and D denote the electrode from/to which an electron tunnels, and the subscripts i and o denote the tunnel-in and tunnel-out processes, respectively. Since there always exists unavoidable electrode- and energy-dependent asymmetry in tunneling probabilities, we have a finite current flow even at $V_{DS} = 0$.[30]

For spectroscopic measurements, we used a Ce@C$_{82}$ SMT itself as a THz detector and performed Fourier transform infrared spectroscopy[23], as shown in Fig. 3a. Figure 3b shows an interferogram of the THz-induced photocurrent measured on a Ce@C$_{82}$ SMT when the gate voltage was set at the peak position of the photocurrent ($V_G$ = -6.82 V; black arrow in Fig. 2b). A clear center peak and an interference pattern can be seen in the interferogram. By calculating its Fourier spectrum, we obtained a THz spectrum for a Ce@C$_{82}$ SMT. As seen in Fig. 3c, two broad peaks with opposite polarities are observed at ~5 meV and ~15 meV. In our previous work,[22] we observed the center-of-mass oscillation of the Ce@C$_{82}$ molecule by using THz bursts generated by femtosecond laser pulses (we replotted the previously obtained THz spectrum for the center-of-mass vibration by a dashed line in Fig. 3c).[22] Since its frequency is too low (~ 500 GHz = ~2 meV) for the measurements with a blackbody light source, the center-of-mass motion was not visible in the present measurements.

To understand the origin of the observed peaks, we have performed transport measurements[29] on a hollow-cage single C$_{84}$ molecule as a reference (see Fig. S2). There are no vibrational modes below 20 meV. In addition, the calculated THz absorption spectrum[31] of an empty C$_{82}$ molecule does not show any vibrational modes below ~25 meV. In contrast, new low-energy excitation modes appear when a metal atom is injected into the hollow C$_{82}$ cage[31,32]. This clearly indicates that the two vibrational modes around 5 meV and 15 meV are associated with the Ce atom in the cage. Indeed, it is predicted that the Ce atom in the C$_{82}$ cage has two vibrational modes[33], as schematically illustrated in the inset of Fig. 3c; the vibrational mode at ~5-6 meV is the bending motion of the Ce atom and the peak at around ~15-16 meV is due to the stretching mode. Previous THz measurements on an ensemble of Ce@C$_{82}$ molecules observed only one broad peak[33] extending from 1 meV to 12 meV and could not resolve the two separate peaks, most likely due to inhomogeneous broadening. The observation of the two rattling modes has become possible only after inhomogeneous broadening has been removed by measuring a single molecule.

Figure 4a plots the interferograms measured at various $V_{DS}$. When $V_{DS}$ is varied even for a small range (-5 mV < $V_{DS}$ < 5 mV), the polarity and the shape of the interferogram change dramatically. Figure 4b shows the obtained photocurrent spectra measured at various $V_{DS}$. We can clearly see two broad photocurrent peaks at ~5 meV and ~15 meV, which are attributed to the bending and stretching modes, respectively.[33]   Figure 4c plots the magnitudes of the two photocurrent peaks as a function of $V_{DS}$.   The magnitude and the polarity of the photocurrent sensitively depend on the bias voltage.   The photocurrents for the two modes at $V_{DS}$ = 0.1 mV have opposite polarities.   When a larger positive or negative bias voltage is applied, however, the two components start to have the same polarities.   As a result, the photocurrent spectra exhibit a rather complicated behavior.   For example, when $V_{DS}$ = 1 mV, the photocurrent peak for the bending mode almost vanishes, while the peak for the stretching mode is suppressed when $V_{DS}$ = -1 mV.   Such a complicated behavior can be understood by considering the fact that the single electron photovoltaic effect arises from a subtle imbalance in the tunnel coupling probabilities of the HOMO and LUMO states with the source and drain electrodes.[30]   Furthermore, note that the photocurrent peaks are very broad; the peak width for the bending mode is ~ 4 meV and that for the stretching mode is even broader (~8 meV).   These broad peak widths are consistent with theoretical predictions that the confinement potential for the Ce atom in the $C_{82}$ cage is very anharmonic[11,12].   In addition, we have found that the spectrum for the stretching mode is broader than that for the bending mode, which is in agreement with theory[34]; it predicted that the curvature of the confinement potential of the Ce atom is more nonparabolic along the radius direction of the cage.[34]

   In summary, we have demonstrated that the THz spectroscopy that uses nanogap electrodes as THz antennas can detect ultrafast rattling motion of a single Ce atom in a $C_{82}$ fullerene cage.   The THz-induced photocurrent spectra exhibited two broad peaks due to the vibron-assisted tunneling process.   The vibrational modes observed at ~5 meV and ~15 meV are respectively attributed to the bending and stretching motions of the Ce atom in the $C_{82}$ cage.   By mapping the THz-induced photocurrent as a function of $V_G$ and $V_{DS}$, we have found that the THz-induced photocurrent originates not only from the photoconductive effect but also from the photovoltaic effect that arises from asymmetry in the tunnel coupling of the HOMO/LUMO with the electrodes.   The present result have demonstrated that the THz spectroscopy have attained an atom-level spatial resolution by using nanogap electrodes.   We believe that such an ultrahigh spatial resolution, together with sub-ps time resolution inherent to THz spectroscopy, will provide a great impact to physics, chemistry, molecular biology, and pharmaceutical sciences.

**Acknowledgements**

We thank Y. Arakawa for discussions on single molecule spectroscopy, C. C. Tang for transport measurements and S. Ishida for his technical support in the sample fabrication process. This work was supported by MEXT KAKENHI on Innovative Areas "Science of hybrid quantum systems" (15H05868), KAKENHI from JSPS (16H06709), and Canon Science Foundation.

**Contributions**

S.Q.D. fabricated the single molecule transistor samples and carried out the terahertz measurements. K.H. conceived and supervised the project. Y.Z. provided assistance in the THz spectroscopy and K.Y. supported the transport measurements. S.Q.D. and K.H. wrote the manuscript with contributions from all authors. All authors contributed to discussions.

**References**

a) Electronic mail: sqdu@iis.u-tokyo.ac.jp

b) Electronic mail: hirakawa@iis.u-tokyo.ac.jp

1    Bergeron, D. E., Castleman, A. W., Morisato, T. & Khanna, S. N. Formation of $Al_{13}I^-$: evidence for the superhalogen character of $Al_{13}$. *Science* **304**, 84 (2004).

2    Khanna, S. N. & Jena, P. Assembling crystals from clusters. *Phys. Rev. Lett.* **69**, 1664-1667 (1992).

3    Feng, M., Zhao, J. & Petek, H. Atomlike, hollow-core–bound molecular orbitals of $C_{60}$. *Science* **320**, 359 (2008).

4    Aoyagi, S. *et al.* A layered ionic crystal of polar $Li@C_{60}$ superatoms. *Nat. Chem.* **2**, 678-683 (2010).

5    Bethune, D. S., Johnson, R. D., Salem, J. R., Devries, M. S. & Yannoni, C. S. Atoms in carbon cages - the structure and properties of endohedral fullerenes. *Nature* **366**, 123-128 (1993).

6    Weiss, F. D., Elkind, J. L., O'Brien, S. C., Curl, R. F. & Smalley, R. E. Photophysics of metal complexes of spheroidal carbon shells. *J. Am. Chem. Soc.* **110**, 4464-4465 (1988).

7    Chai, Y. *et al.* Fullerenes with metals inside. *J. Phys. Chem.* **95**, 7564-7568 (1991).

8    Takata, M. *et al.* Confirmation by X-ray diffraction of the endohedral nature of the metallofullerene $Y@C_{82}$. *Nature* **377**, 46 (1995).

9    Muthukumar, K. *et al.* Endohedral fullerene $Ce@C_{82}$ on Cu(111): orientation, electronic structure, and electron-vibration coupling. *J. Phys. Chem. C* **117**, 1656-1662 (2013).

10   Suenaga, K. *et al.* Visualizing and identifying single atoms using electron energy-loss spectroscopy with low accelerating voltage. *Nat. Chem.* **1**, 415 (2009).

11   Andreoni, W. & Curioni, A. Freedom and constraints of a metal atom encapsulated in fullerene cages. *Phys. Rev. Lett.* **77**, 834-837 (1996).

12   Jorn, R., Zhao, J., Petek, H. & Seideman, T. Current-driven dynamics in molecular junctions: endohedral fullerenes. *ACS Nano* **5**, 7858-7865 (2011).

13   Cocker, T. L., Peller, D., Yu, P., Repp, J. & Huber, R. Tracking the ultrafast motion of a single molecule by femtosecond orbital imaging. *Nature* **539**, 263 (2016).

14  Jelic, V. *et al.* Ultrafast terahertz control of extreme tunnel currents through single atoms on a silicon surface. *Nat. Phys* **13**, 591 (2017).

15  Imada, H. *et al.* Single-molecule investigation of energy dynamics in a coupled plasmon-exciton system. *Phys. Rev. Lett.* **119**, 013901 (2017).

16  Park, H. *et al.* Nanomechanical oscillations in a single-$C_{60}$ transistor. *Nature* **407**, 57-60 (2000).

17  Kubatkin, S. *et al.* Single-electron transistor of a single organic molecule with access to several redox states. *Nature* **425**, 698-701 (2003).

18  Osorio, E. A. *et al.* Electronic excitations of a single molecule contacted in a three-terminal configuration. *Nano Lett.* **7**, 3336-3342 (2007).

19  de Leon, N. P., Liang, W., Gu, Q. & Park, H. Vibrational excitation in single-molecule transistors: deviation from the simple Franck−Condon prediction. *Nano Lett.* **8**, 2963-2967 (2008).

20  Clemmer, D. E., Hunter, J. M., Shelimov, K. B. & Jarrold, M. F. Physical and chemical evidence for metallofullerenes with metal atoms as part of the cage. *Nature* **372**, 248-250 (1994).

21  Yoshida, K., Shibata, K. & Hirakawa, K. Terahertz field enhancement and photon-assisted tunneling in single-molecule transistors. *Phys. Rev. Lett.* **115**, 138302 (2015).

22  Du, S., Yoshida, K., Zhang, Y., Hamada, I. & Hirakawa, K. Terahertz dynamics of electron–vibron coupling in single molecules with tunable electrostatic potential. *Nat. Photon.* **12**, 608-612 (2018).

23  Zhang, Y. *et al.* Terahertz intersublevel transitions in single self-assembled InAs quantum dots with variable electron numbers. *Nano Lett.* **15**, 1166-1170 (2015).

24  Park, H., Lim, A. K. L., Alivisatos, A. P., Park, J. & McEuen, P. L. Fabrication of metallic electrodes with nanometer separation by electromigration. *Appl. Phys. Lett.* **75**, 301-303 (1999).

25  Umeno, A. & Hirakawa, K. Nonthermal origin of electromigration at gold nanojunctions in the ballistic regime. *Appl. Phys. Lett.* **94**, 162103 (2009).

26  Strachan, D. *et al.* Controlled fabrication of nanogaps in ambient environment for molecular electronics. *Appl. Phys. Lett.* **86**, 043109 (2005).

27  Ward, D. R., Huser, F., Pauly, F., Cuevas, J. C. & Natelson, D. Optical rectification and field enhancement in a plasmonic nanogap. *Nat. Nano.* **5**, 732-736 (2010).

28  Seo, M. A. *et al.* Terahertz field enhancement by a metallic nano slit operating beyond the skin-depth limit. *Nat. Photon.* **3**, 152-156 (2009).

29  Okamura, N., Yoshida, K., Sakata, S. & Hirakawa, K. Electron transport in endohedral metallofullerene Ce@$C_{82}$ single-molecule transistors. *Appl. Phys. Lett.* **106**, 043108 (2015).

30  Zhang, Y. *et al.* Gate-controlled terahertz single electron photovoltaic effect in self-assembled InAs quantum dots. *Appl. Phys. Lett.* **107**, 103103 (2015).

31  Andreoni, W. & Curioni, A. Ab initio approach to the structure and dynamics of metallofullerenes. *Appl. Phys. A* **66**, 299-306 (1998).

32  Muthukumar, K. & Larsson, J. A. A density functional study of Ce@$C_{82}$: explanation of the Ce preferential bonding site. *J. Phys. Chem. A* **112**, 1071-1075 (2008).

33  Lebedkin, S., Renker, B., Heid, R., Schober, H. & Rietschel, H. A spectroscopic study of M@$C_{82}$ metallofullerenes: Raman, far-infrared, and neutron scattering results. *Appl. Phys. A* **66**, 273-280 (1998).

34  Zheng, W., Ren, S., Tian, D. & Hao, C. The dynamic motion of a M (M = Ca, Yb) atom inside the $C_{74}$ ($D_{3h}$) cage: a relativistic DFT study. *J. Mol. Model.* **19**, 4521-4527 (2013).

**Figure Captions**

**Figure 1  Single Ce@C$_{82}$ molecule transistor sample with terahertz antenna electrodes.**  **a**, Schematic of a single Ce@C$_{82}$ molecule transistor structure.  A single Ce@C$_{82}$ molecule is captured in a sub-nm gap of metal electrodes.  **b**, SEM image of the nanojunction region of a single molecule transistor (SMT).  The structure consists of three electrodes; the source (Au), drain (Au), and gate (NiCr).  An insulating Al$_2$O$_3$ layer is deposited between the gate and the source/drain.  After the electrical break-junction process, the gap size in the order of ~1 nm is usually obtained.  **c**, Resonance spectrum of a designed bow-tie antenna calculated by using the finite element method (FEM).  A broad resonance peak appears around 10 meV (~2.5 THz).  The inset shows the geometry of the bow-tie antenna.  The color shows the simulated electric-field distribution at the resonant frequency (10 meV).

**Figure 2  Vibron-assisted tunneling due to ultrafast motion of the Ce atom.**  **a**, Coulomb stability diagram of a Ce@C$_{82}$ SMT.  White dashed lines are eyeguides for the boundaries of the Coulomb diamonds.  Black dashed lines are fits to the vibrational excited state lines.  **b**, The single electron tunneling current $I_{DS}$ (black) and the THz-induced photocurrent $I_P$ (red) measured as a function of $V_G$ at $V_{DS}$ = 0.1 mV.  $I_P$ is large near the Coulomb peak and decreases as $V_G$ approaches -6 V (see the blue dotted line).  **c & d**, THz-induced vibron-assisted tunneling process when $V_G$ is on the left side of the charge degeneracy point.  **e & f**, THz-induced vibron-assisted tunneling process when $V_G$ is on the right side of the charge degeneracy point.  $\Gamma_{Si}$: tunnel-in rate from the source; $\Gamma_{Di}$: tunnel-in rate from the drain;  $\Gamma_{So}$: tunnel-out rate to the source;  $\Gamma_{Do}$: tunnel-out rate to the drain.

**Figure 3  THz spectroscopy of ultrafast single atom motion inside the fullerene cage.**  **a**, Schematic illustration of the experimental setup for the Fourier transform infrared spectroscopy. Broadband THz radiation (blackbody light source at ~1100 K) is focused onto a SMT mounted in a $^4$He cryostat.  **b**, Autocorrelation trace of the THz-induced photocurrent measured at $V_G$ indicated by the black arrow in **Fig. 2b** ($V_G$ = -6.82 V).  **c**, Spectrum of the THz-induced photocurrent obtained

by Fourier transformation of the interferogram shown in **b**. Two broad peaks at around ~5 meV and ~15 meV with opposite polarities are observed. The peak around 5 meV is due to the bending motion of the encapsulated Ce atom and the one around 15 meV is due to the stretching motion. The dashed line is the spectrum of the center of mass oscillation of the whole molecule measured by the time-domain THz spectroscopy (replotted from ref. 22).

**Figure 4 Bias-dependence of the THz photocurrent spectra due to ultrafast motion of a trapped single Ce atom.** **a**, Bias-dependence of the interferograms of the THz-induced photocurrent measured on a single Ce@$C_{82}$ molecule measured at $V_G$ = -6.82 V. **b**, The corresponding THz spectra of **a**. **c**, The magnitudes of the two photocurrent peaks as a function of $V_{DS}$ (- 5 mV < $V_{DS}$ < 5 mV).

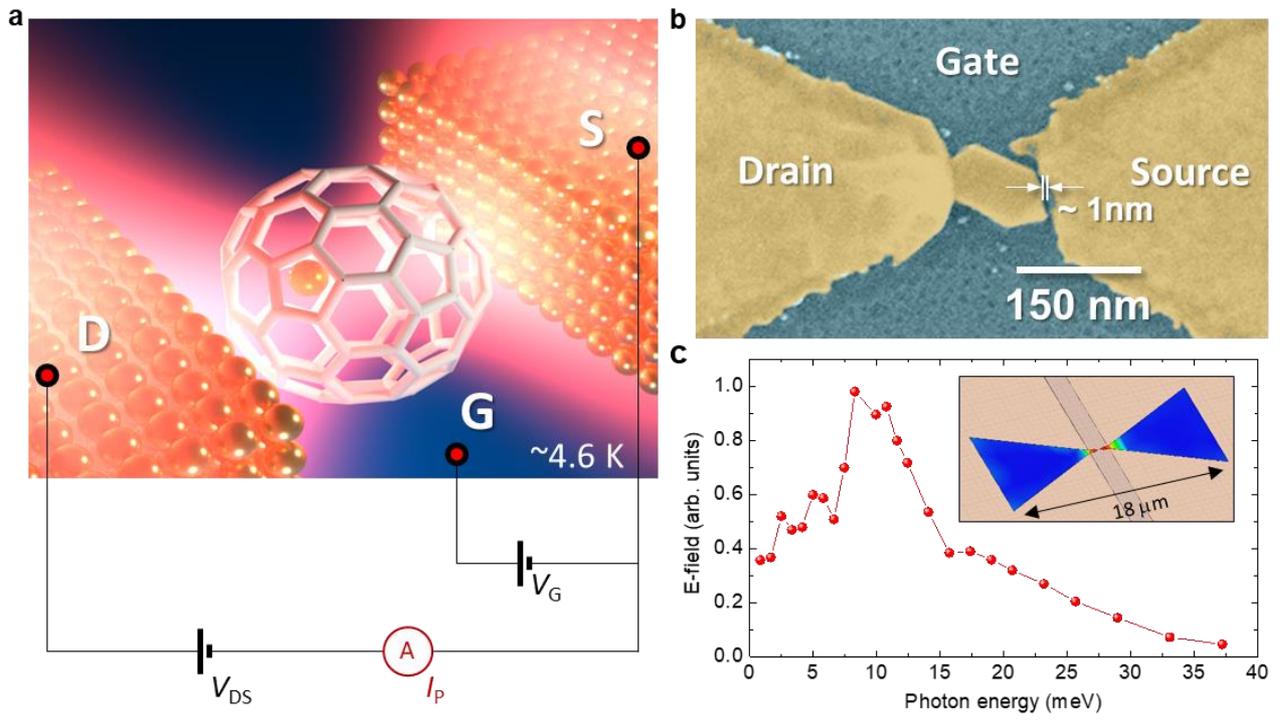

Figure 1 S. Q. Du, et al.

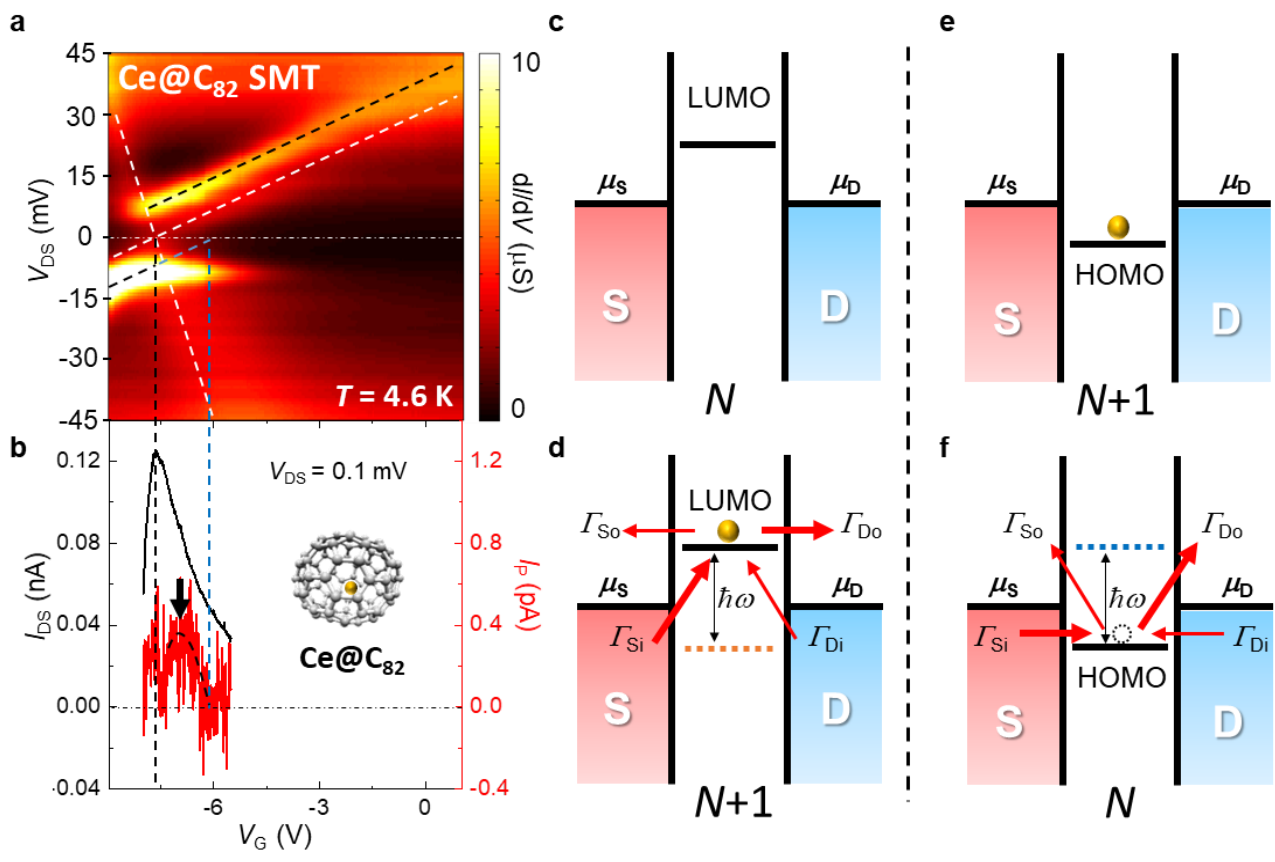

Figure 2 S. Q. Du, et al.

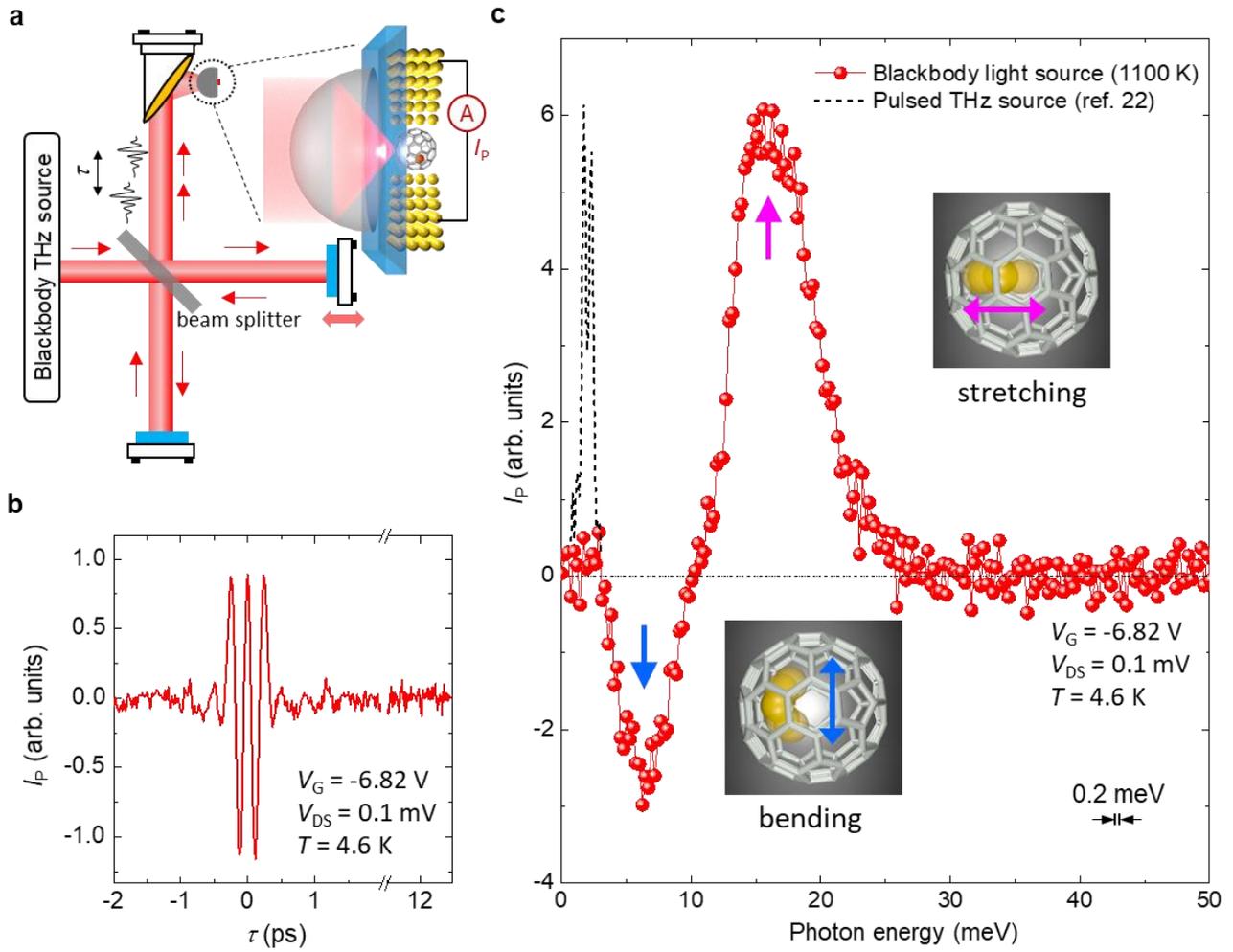

Figure 3 S. Q. Du, et al.

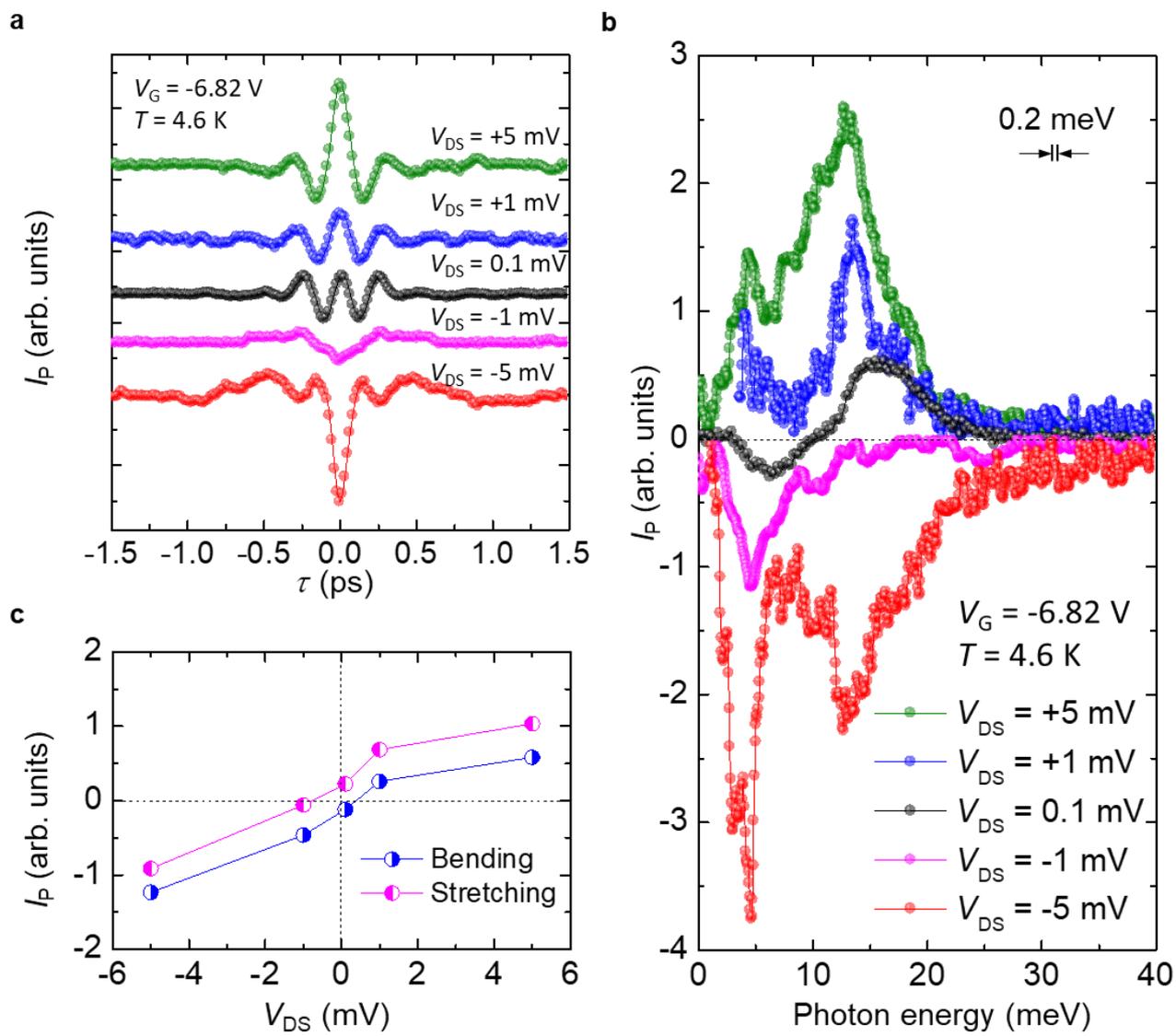

Figure 4 S. Q. Du, et al.